\documentclass[12pt,titlepage]{article}
\usepackage{doublespace,fullpage,spacing}

\newcommand{\al}{\alpha}
\newcommand{\di}{\partial}
\newcommand{\benonumber}{\begin{displaymath}}
\newcommand{\eenonumber}{\end{displaymath}}
\newcommand{\be}{\begin{equation}}
\newcommand{\ee}{\end{equation}}
\newcommand{\eref}[1]{(\ref{#1})}

\newcommand{\ep}{\varepsilon}
\newcommand{\kk}{\kappa}
\newcommand{\half}{\frac{1}{2}}
\newcommand{\third}{\frac{1}{3}}

\newcommand{\sixth}{\frac{1}{6}}

\newcounter{saveeqn}%
\newcommand{\alpheqn}{\setcounter{saveeqn}{\value{equation}}%
\stepcounter{saveeqn}\setcounter{equation}{0}%
\renewcommand{\theequation}
	{\mbox{\arabic{saveeqn}\alph{equation}}}}%
\newcommand{\reseteqn}{\setcounter{equation}{\value{saveeqn}}%
\renewcommand{\theequation}{\arabic{equation}}}%

\newcommand{\mainlabel}[1]{\renewcommand{\theequation}{\arabic{saveeqn}}%
\label{#1}
\renewcommand{\theequation}{\mbox{\arabic{saveeqn}\alpha{equation}}}}%

\title{Possible Wormhole Solutions in (4+1) Gravity}
	
\author{A. G. Agnese\thanks{Dipartimento di Fisica, Universit\`{a} di Genova,
Instituto Nazionale di Fisica Nucleare, Sezione di Genova, Italy} 
\and A. P. Billyard\thanks{Dept. Physics, Dalhousie Univ., Halifax, NS, 
Canda} 
\and H. Liu\thanks{Dept. Physics, Univ. Waterloo, Waterloo, ON, Canda} 
\and P. S. Wesson \thanks{Dept. Physics, Univ. Waterloo, 
Waterloo, ON, Canada}}

\begin{document}
\maketitle

\begin{abstract}
We extend previous analyses of soliton solutions in $(4+1)$ gravity to
new ranges of their defining parameters.  The geometry, as studied using
invariants, has the topology of wormholes found in $(3+1)$ gravity.  In the
induced-matter picture, the fluid does not satisfy the strong energy 
conditions, but its gravitational mass is positive.  We infer the possible
existance of $(4+1)$ wormholes which, compared to their $(3+1)$ counterparts, are less
exotic.
\end{abstract}

\section{Introduction}

Multi-dimensional theories of gravity have been studied in great
detail since 1921 \cite{Kaluza1921, Klein1926a,Klein1926b}.  For
extensive reviews of these theories, we refer the reader to papers by
Duff \cite{Duff1994} and by Overduin and Wesson \cite{Overduin1997}.
In five dimensions, there is considerable literature, both in
cosmology and astrophysics.  Specific to the latter there is a class
of static, spherically symmetric solutions which are parameterized by
three constants: $M$, $\epsilon$, $\kappa$, (the second two of which
are related to one another).  These solutions are the analogues to the
four-dimensional Schwarzschild solution, allowed by the non-applicability of
Birkhoff's theorem in $(4+1)$ dimensions.  These ``soliton'' solutions
were first extensively studied by Gross and Perry \cite{Gross1983}
and by Davidson and Owen \cite{Davidson1985}, although the solutions
were known previous to them.  These solutions have been studied
within the induced-matter theory of Wesson \cite{Wesson1992d,Wesson1994}, 
in which the
five-dimensional vacuum solutions give rise to four-dimensional
static, spherically symmetric solutions with a radiation-like matter
field.

In a recent paper \cite{Billyard1995}, the definition of the
(three-dimensional) spatial origin to these solitons was discussed,
as were the definitions of mass.  The assumptions
$\epsilon>0$ and $\kappa>0$ were used in order for the induced
four-dimensional gravitational mass, pressure and density to be
positive.  Below, we relax previous assumptions and consider $\epsilon<0$ 
and $\kappa<0$, finding that while the induced matter does not satisfy the
strong energy condition, its gravitational mass is still positive.  We 
also study the geometry of the $(4+1)$ spacetime using invariants, finding
it has the topology typical of wormholes \cite{Morris1988a, Morris1988b, Visser1995}.  We
infer the possible existance of $(4+1)$ wormholes which, compared to their
$(3+1)$ counterparts, are in some ways less ``exotic''.

\section{Five-Dimensional Soliton Solutions}

Following Davidson and Owen's notation \cite{Davidson1985}, the line element 
of the solitons can be written
\be   
ds^2=T^2(\rho) dt^2 - S^2(\rho) (d\rho^2 + \rho^2 d\Omega^2) - \Phi^2(\rho) 
d\psi^2,
\label{isotropic}                                                  
\ee
where \alpheqn
\begin{eqnarray}
\mainlabel{coeff1}
T(\rho) & = & \left( \frac{\rho-M/2}{\rho+M/2} 
		  \right)^{\epsilon k}, \label{A1} \\
S(\rho) & = & \left( 1-\frac{M^2}{4\rho^2} \right)
	   \left( \frac{\rho+M/2}{\rho-M/2} 
		  \right)^{\epsilon(k-1)}, \label{B1} \\
\Phi(\rho) & = & \left( \frac{\rho+M/2}{\rho-M/2} 
		  \right)^{\epsilon}. \label{C1}
\end{eqnarray}\reseteqn
Here $d\Omega \equiv d\theta^2+\sin^2\!\theta d\phi^2$ as usual, $M$ is a 
constant with units of mass, and we set $c=G=1$.  The two constants 
$\epsilon$ and $\kappa$ are related by $\epsilon=\pm 1/\sqrt{\kappa^2-
\kappa +1}$, but we leave both explicit for algebraic convenience.  In 
\cite{Billyard1995} it was 
shown that for $0<\kappa<\infty$ and $\epsilon>0$ these solutions represent
naked singularities with an origin located at $\rho=\half M$.  In the limiting
case $(\epsilon,\kappa,\epsilon\kappa) \rightarrow (0,\infty,1)$, which 
is the ``Schwarzshild'' limit, a black hole is obtained with its 
origin located at $\rho=- \half M$. These results were obtained from examining 
at which radii do surface areas vanish, and from examining divergences in 
the Kretchmann scalar
\begin{eqnarray}
 \nonumber
 R_{ABCD}R^{ABCD} & = & \left[\rho^4 
    - 2\ep (\kk -1)(2+\ep ^2\kk) \rho^3\left(\frac{M}{2}\right) 
     + 2(3-\ep ^4\kk ^2)\rho^2\left(\frac{M}{2}\right)^2 \right.\\  & &
	\left.-2\ep (\kk -1)(2+\ep ^2\kk ) \rho\left(\frac{M}{2}\right)^3 
	+ \left(\frac{M}{2}\right)^4\right]
  \frac{48 M^2\rho^6}{\left(\rho^2-\frac{M^2}{4}\right)^8}
   \left(\frac{\rho-\frac{M}{2}}{\rho+\frac{M}{2}} \right)^{4\ep(\kk-1)}.
 \label{Kretchmann}
\end{eqnarray} 
Here, large Latin indices run $01234$, and below we will use small Greek
indices that run $0123$.

In Billyard, Kalligas and Wesson \cite{Billyard1995} a restriction of 
$\epsilon>0$ and $\kappa>0$ was made based on induced-matter arguments.  
However, the gravitational and inertial mass (defined in five dimensions) 
are
\begin{eqnarray*}
M_{grav}&=&\epsilon\kappa M, \\
M_{inert}&=&\epsilon(\kappa-\half)M.
\end{eqnarray*}
These only require $\epsilon\kappa>0$ and $\epsilon(\kappa-1/2)>0$.  For 
$\epsilon>0$, this sets $\kappa>\half$, but for $\epsilon<0$ one may have $k<0$
and both masses will still remain positive.  In fact, the only qualitative 
difference the latter makes to \eref{isotropic} is that $\Phi(\rho)$ is 
inverted.  It is this latter range for $(\epsilon,\kappa)$ which was not 
previously considered and which leads to possible wormhole solutions.  

The existence of these is not apparent in the $\rho$coordinates of 
\eref{isotropic}, but will become so if we consider an $r-$coordinate give
by
\be
r = \rho \left( 1-\frac{M^2}{4\rho^2}\right) \left( 
    \frac{\rho+M/2}{\rho-M/2}\right) ^{\epsilon(k-1)}. \label{trfm}
\ee 
The nature of this transformation is shown in Figure 1.
The curve which ends at $\rho
=\half M$ is representative of the transformation for $\epsilon>0$, 
$\kappa >0$ for which $-1\leq \epsilon(\kappa-1) < 1$.  The curve which 
diverges at $\rho=0$ is the limiting Scwharzschild curve.  The solid curve
represents the transformation \eref{trfm} for which $\epsilon<0$, $\kappa<0$
such that $1<\epsilon(\kappa-1)<2/\sqrt{3}$.  In this regime, the minimum of
$r$ is located at ($\rho_T$,$r_T$) and is at
\alpheqn\begin{eqnarray}
\rho_T & = & \half M\left[\epsilon(\kappa-1)+\sqrt{\epsilon^2(\kappa-1)^2
- 1}\right], \\
r_T & = &M\frac{\left[ \epsilon(\kappa+1)+1 \right]^{\half\epsilon(\kappa-1)
   +\half}} {\left[ \epsilon(\kappa+1)-1 \right]^{\half\epsilon(\kappa-1)
   -\half}}.
\end{eqnarray}\reseteqn
The location of this minimum ranges from $\rho_T\approx 0.9 M, r_T\approx 
2.6M$ for $\kappa=-1/2$, to $(\half M,2M)$ for $\kappa \rightarrow -\infty$.  
Under transformation \eref{trfm}, the isotropic spacetime of \eref{isotropic} 
becomes
\be   
ds^2=A^{\epsilon\kappa} dt^2 - \frac{A}{B^2}dr^2 - r^2 d\Omega^2 - 
	  A^{-\epsilon} d\psi^2, \label{curvature}
\ee
where \alpheqn
\begin{eqnarray}
\mainlabel{coeff2}
A & = & 1-\frac{2M}{R} \label{A2}, \\
B & = & 1- \frac{\left[ 1+\epsilon(k-1) \right] M}{R}, \label{B2}  \\
R & \equiv &  \rho\left(1+\frac{M}{2\rho} \right)^2.
\end{eqnarray}\reseteqn
These ``quasi-curvature'' coordinates are useful in calculations such as 
surface areas of spheres at fixed radii: e.g., $\mathcal A \equiv 4\pi r^2$.  It is 
apparent from \eref{curvature} that there is a divergence in the metric 
(the $g_{rr}$ term) at $R=[\epsilon(\kappa-1)+1] M$, where 
$\rho=\rho_T$ and $r=r_T$.  This is a coordinate artifact;
the other metric components neither diverge nor vanish at this point, and 
the Kretchmann scalar \eref{Kretchmann} remains well behaved.  
This suggests that this is the throat of a wormhole.  If so, an observer 
travelling from $r>r_T$ towards the throat would reach $r=r_T$ in a finite 
time ($g_{tt}$ is well defined) and would then proceed to travel in $r>r_T$ in 
the other spacetime.  The surface area of this throat is determined by the
values of $\epsilon$ and $\kappa$, namely,
\be
\mathcal A= 4\pi M^2\frac{\left[ \epsilon(\kappa+1)+1 \right]^{\epsilon(\kappa-1)+1}}{\left[ \epsilon(\kappa+1)-1 \right]^{\epsilon(\kappa-1)-1}}.
\ee

For $-\infty<\kappa <0$, there is still a curvature singularity
present at $\rho=\half M$, so only one of the spacetimes is
asymptotically flat ($\rho>\rho_T$), whilst the other has a curvature
singularity at $r=\infty$, ($\rho=\half M$).

\section{Induced Matter}

In the induced-matter approach to $(4+1)$ gravity 
\cite{Overduin1997,Wesson1992d,Wesson1994,Billyard1995}, the field equations in terms of the five-dimensional Ricci tensor are
\be
R_{AB}=0, \label{fiveRicci}
\ee
but the first ten components are rewritten as the four-dimensional Einstein
equations with an effective or induced energy-momentum tensor given by
\begin{eqnarray}
\nonumber
8\pi T_{\alpha \beta} & =  & 
	\frac{\Phi_{\al;\beta}}{\Phi} - \frac{\pm 1}{2 \Phi^2} \left\{
\frac{\stackrel{*}{\Phi}\stackrel{*}{g_{\al\beta}} }{\Phi} 
- \stackrel{**}{g_{\al\beta}} + g^{\mu \nu} \stackrel{*}{g_{\al 
\mu}} \stackrel{*}{g_{\beta\nu}}\right. \\
	 & & \left. - \frac{g^{\mu\nu}\stackrel{*}{g_{\mu\nu}} 
\stackrel{*}{g_{\al\beta}}}{2}+\frac{g_{\al\beta}}{4} \left[ 
\stackrel{*}{g^{\mu\nu}} \stackrel{*}{g_{\mu\nu}} + \left( g^{\mu\nu} 
\stackrel{*}{g_{\mu\nu}} \right) ^2 \right] \right\}. \label{Tmn}
\end{eqnarray}
Here the extra metric component is $g_{44}= g_{\psi \psi}=\pm\Phi^2$
($g_{4\alpha}=0$), a semicolon represents the usual (3+1) covariant
derivative, $\Phi_{\alpha}\equiv\di\Phi/\di x^\alpha$ and an overstar
represents $\di/\di\psi$.  The soliton solutions \eref{isotropic},
\eref{coeff1} satisfy \eref{fiveRicci}, and its associated fluid has
matter properties given by \eref{Tmn}.  For \eref{isotropic} it was
found \cite{Wesson1992d} that the induced matter was a bath of
radiation, satisfying $\mu=3 p$, where $\mu$ is the energy density and
$p$ is the average pressure.

As previously mentioned, the assumption of $\epsilon>0$, $\kappa>0$
was used because in the induced-matter scenario the effective
four-dimensional mass-energy density, averaged pressure and induced
gravitational mass are respectively\cite{Wesson1992d,Wesson1994},
\alpheqn
\begin{eqnarray}
8\pi \mu & = &  \frac{\epsilon^2 \kappa M^2}{\rho^4 \left(1-\frac{M^2}{4\rho^2} 
\right)^4}\left( \frac{\rho-\frac{M}{2}}{\rho+\frac{M}{2}}\right)^
{2\epsilon(\kappa-1)}, \\
8\pi p & = & \third \mu, \label{pave} \\
M_{grav} & = & \epsilon\kappa\left(\frac{\rho-\frac{M}{2}}
    {\rho+\frac{M}{2}}\right)^\epsilon M.
\end{eqnarray} \reseteqn
Clearly, the strong energy condition, $\mu+3p>0$, is satisfied only
for $\kappa>0$ and so one further assumes $\epsilon>0$ so that
$M_{grav}>0$.  However, even if $\kappa<0$ and $\epsilon<0$ so that
the induced matter does not satisfy the strong energy condition, the
induced gravitational mass is {\em still} positive.  Violation of the
strong energy condition is not new, and is always encountered when
considering wormholes in conventional four-dimensional theories (see
for example \cite{Morris1988a, Morris1988b,Visser1995}).

With the induced matter, we are in a position to calculate the tension
of the wormhole's throat, which is the negative value of the radial
pressure \cite{Morris1988b}.  The pressure \eref{pave} is
obtained from the average of the three pressures
\alpheqn\begin{eqnarray}
8\pi p_r & \equiv & 8\pi T^r_r = \frac{8\pi\mu}{\epsilon\kappa} 
     \left[\frac{M}{2\rho}\right]\left[\left(\frac{2\rho}{M}\right)^2
      -\epsilon(\kappa-2)\frac{2\rho}{M}+1\right], \\
8\pi p_\theta & \equiv & 8\pi T^\theta_\theta = -\frac{8\pi\mu}{2\epsilon\kappa}
     \left[\frac{M}{2\rho}\right]\left[ \left(\frac{2\rho}{M}\right)^2
     -2\epsilon(\kappa-1)\frac{2\rho}{M}+1\right], \\
8\pi p_\phi & = & 8 \pi p_\theta.
\end{eqnarray}\reseteqn
Now, for $\kappa<0$, $p_r$ remains negative througout $\half M < \rho
< \infty$, whereas the transverse pressures $p_\theta$ and $p_\phi$
are negative for $\rho<\rho_T$, zero at the throat, and positive for
$\rho>\rho_T$.  Through some algebra, it may be verified that at the
throat, the tension is
\be
\tau \equiv -p_r = -\mu > 0,
\ee
where $0<\epsilon\kappa<1$.  We conclude this calculation with the note that 
regular four-dimensional wormholes typically have $\tau>\mu$ \cite{Morris1988b}, 
which is not found here.

Whilst we are considering the induced matter of \eref{isotropic}, let
us examine the scalars $I_1\equiv R^\alpha_\alpha$, $I_2\equiv
R^{\alpha \beta} R_{\alpha \beta}$ and $I_3\equiv R^{\alpha \beta
\gamma\delta} R_{\alpha \beta \gamma\delta}$ on the four-dimensional
hypersurfaces where the induced matter is defined.  This will enable
us to ascertain where four-dimensional singularities occur.  Because
the induced matter is radiation-like, then $I_1\equiv 0$.  The other two 
scalars are given by
\begin{eqnarray}
I_2 & = & \frac{6\epsilon^2 M^2}{R^6} A^{2[\epsilon(\kappa-1)-1]} \left[
	    1 + \frac{D_1 M}{R A} + \frac{D_2 M^2}{R^2 A^2} \right], \\
I_3 & = & \frac{48 M^2}{R^6} A^{2[\epsilon(\kappa-1)-1]} \left[
	    D_3 + \frac{D_4 M}{R A} + \frac{D_5 M^2}{R^2 A^2} \right],
\end{eqnarray}
where
\alpheqn\begin{eqnarray}
D_1 &=& 2-\frac{2}{3}\epsilon\kappa+3\epsilon, \\
D_2 &=& \sixth[\epsilon^2\kappa^2 +2(\epsilon(k-1)-1)^2+(\epsilon(k-2)-2)^2],
 \\
D_3 &=& 1-\half\epsilon^2, \\
D_4 &=& -\third[a^2(\epsilon(3\kappa-2)-3) +(\epsilon(\kappa-1)-1)^3-
  (\epsilon(\kappa-1)-1)], \\
D_5 &=& \frac{1}{12}[a^2(\epsilon(2\kappa-1)-2)^2 +(\epsilon(\kappa-1)-1)^4+
		 2(\epsilon^2\kappa^2+1)(\epsilon(\kappa-1)-1)^2]. 
\end{eqnarray}\reseteqn
For the range $\epsilon<0$ and $\kappa<0$, excluding the Schwarzschild limit,
it may be easily verified that these scalars are divergent at $\rho=\half M$
and are well defined at the throat of the wormhole, $\rho=\rho_T$.

\section{Discussion and Conclusion}

By relaxing restrictions on the parameters $\epsilon$ and $\kappa$, we
find that the static, spherically symmetric solutions in five
dimensions can be interpreted as wormholes.  That is, there are
solutions where there is a bridge between two spacetimes; one is
asymptotically flat and the other containing a curvature singularity
at spatial infinity.  In the induced matter scenario, the induced
mass-energy density and pressure violate the strong energy condition,
as do the matter sources for wormholes in four dimensions, yet the
gravitational mass remains positive.  However, unlike the solitons
(i.e., equation \eref{isotropic} with $\epsilon>0,\kappa>0$) this
induced gravitational mass diverges in the limit $\rho\rightarrow
\half M$ ($r\rightarrow\infty$), and so may be considered the source
of the singularity found there.  We were able to calculate the tension
in the throat and, unlike four-dimensional wormholes, the magnitude of
the tension is less than that of the mass-energy density.  In the
asymptotically flat universe, the transverse (angular) pressures are
positive while they are negative in the asymptotically singular
spacetime.  In both space-times, the radial pressure is negative.  We
can envision an observer travelling from the asymptotically flat
space-time into the asymptotically singular space-time and encountering
shells of matter (radiation or ultra-relativistic particles) whose
density and pressure steadily increase with radial distance and
eventually diverge at infinity.  Although the matter seems exotic in
four dimensions, the five-dimensional spacetime (on both sides of the
throat) is that of a vacuum and so the energy conditions are
(trivially) satisfied.  Therefore, we find that Kaluza-Klein theory in
the context of induced matter theory can help alleviate concerns about
the existance of ``exotic'' matter \cite{Morris1988a} in four
dimensions; the matter observed in four dimensions is indeed derived
from a five-dimensional theory where the energy conditions {\em are}
satisfied.

In the limiting Schwarzschild case, we note in passing that the
wormhole becomes a black hole once again, due to the vanishing
component of $g_{tt}$; the time it would take for an infalling object
to reach the throat, as measured by an observer in the asymptotically
flat spacetime, would be infinite.  Within the wormhole class,
however, the time would remain finite (i.e., $g_{tt}(\rho_T)\neq 0$).

As is quite evident, the parameter $\kappa$ crucially determines what
type of manifold is described by \eref{isotropic}: soliton
($\kappa<0$), wormhole ($\kappa>0$) or black hole ($\kappa\rightarrow
\pm \infty$), and we note here the physical interpretation of this
constant.  Because of the definition of the gravitational mass,
$M_{grav}=\epsilon\kappa M$ (this is defined at large $\rho$, see
\cite{Billyard1995}), then $\kappa$ is a function of the ration of the
gravitational mass to the constant $M$.  However, there is also
another interpretation.  As was recently examined, there is a formal
equivalence between a five-dimensional vacuum solution which is
independent of the extra coordinate and a four-dimensional vacuum
theory of gravity minimally coupled to a scalar field, $\varphi$ (and
also to a four-dimensional vacuum scalar-tensor theory)
\cite{Billyard1997}.  In this formal equivalence, the scalar field
goes as $\varphi \propto \ln(g_{44})$ and so, the constant $\epsilon$
is this field's strength.  Following the syntax of Agnese and La
Camera \cite{Agnese1985}, the field's strength is given by $\sigma/M$,
where $\sigma$ is a scalar charge (constant).  For the wormhole
solutions, the scalar charge is negative whilst for the soliton
solutions it is positive.  Because of the constraint
$\epsilon^2(\kappa^2-\kappa+1)=1$, this scalar charge constant is
related to the gravitational mass constant and to the constant $M$ by
\benonumber
M^2 = M_{grav}^2-\sigma M_{grav} + \sigma^2.
\eenonumber
Finally, because of this formal equivalence, there are similar wormhole
solutions both in vacuum theories of general relativity coupled to a scalar 
field and in vacuum scalar-tensor theories, the transformations between all 
three are not singular at the throat of these wormholes.  However, it should
be stressed that although there is a mathematical equivalence, the theories
are distinct from each other physically, and the wormholes discussed here 
arise from the non-trivial curvature of the extra dimension, the matter 
properties are geometric in origin.

\noindent{\bf Acknowledgments}
This work was funded by Instituto Nazionale di Fisica Nucleare and the 
Natural Sciences and Engineering Research Council.

\noindent  Figure 1: Transformation between quasi-curvature and isotropic 
co-ordinates


\begin{thebibliography}{10}

\bibitem{Kaluza1921}
T.~H. Kaluza,
\newblock Per. Akad. Phys. Math. K1 {\bf 54}, 966 (1921).

\bibitem{Klein1926a}
O.~Klein,
\newblock Nature {\bf 118}, 516 (1926).

\bibitem{Klein1926b}
O.~Klein,
\newblock Z. Phys. {\bf 37}, 875 (1926).

\bibitem{Duff1994}
J.~Duff,
\newblock hep-th/9410046,
\newblock preprint.

\bibitem{Overduin1997}
J.~Overduin and P.~S. Wesson,
\newblock Phys. Rep. {\bf 283}, 303 (1997).

\bibitem{Gross1983}
D.~J. Gross and M.~J. Perry,
\newblock Nucl. Phys. B {\bf 226}, 29 (1983).

\bibitem{Davidson1985}
A.~Davidson and D.~A. Owen,
\newblock Phys. Letts. B {\bf 155}, 247 (1985).

\bibitem{Wesson1992d}
P.~S. Wesson,
\newblock Phys. Lett. B {\bf 276}, 299 (1992).

\bibitem{Wesson1994}
P.~S. Wesson and J.~P. de~Leon,
\newblock Class. Quant. Grav. {\bf 11}, 1341 (1994).

\bibitem{Billyard1995}
A.~P. Billyard, D.~Kalligas, and P.~S. Wesson,
\newblock Int. J. Mod. Phys. D {\bf 4}, 639 (1995).

\bibitem{Morris1988a}
M.~S. Morris, K.~S. Thorne, and U.~Yrtsever,
\newblock Phys. Rev. Lett. {\bf 61}, 1446 (1988).

\bibitem{Morris1988b}
M.~S. Morris and K.~S. Thorne,
\newblock American J. Phys. {\bf 56}, 395 (1988).

\bibitem{Visser1995}
M.~Visser,
\newblock {\em Lorentzian Wormholes: From {E}instein to {H}awking},
\newblock American Institute of Physics, Woodbury, New York, 1995.

\bibitem{Billyard1997}
A.~P. Billyard and A.~A. Coley,
\newblock Mod. Phys. Lett. A {\bf 12}, 2121 (1997).

\bibitem{Agnese1985}
A.~G. Agnese and M.~L. Camera,
\newblock Phys. Rev. D {\bf 31}, 1280 (1985).

\end{thebibliography}
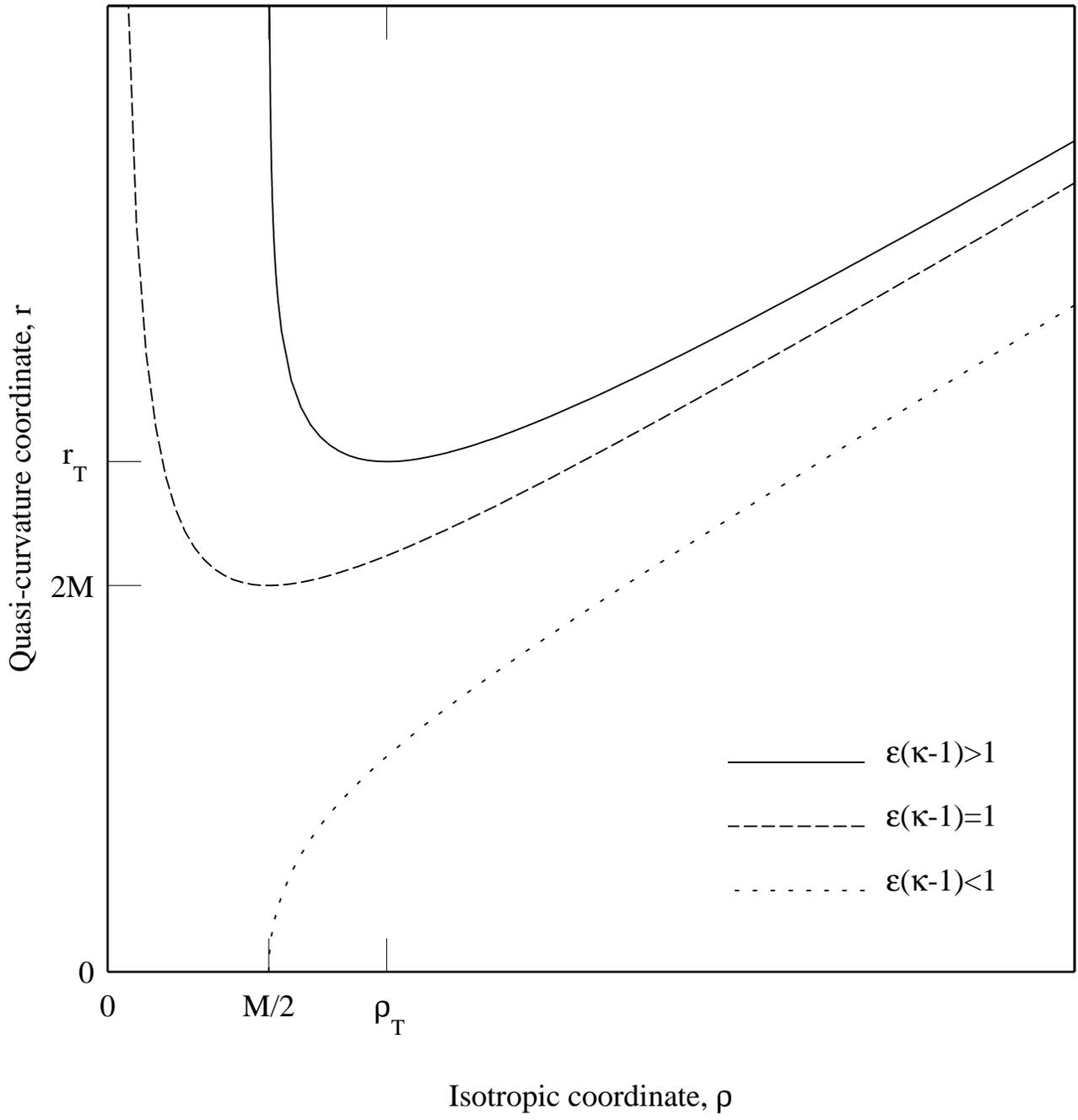
\end{document}